\documentclass[letter]{aa}

\usepackage{graphicx}
\usepackage{txfonts}
\usepackage{natbib}
\usepackage{siunitx}
\usepackage{multirow}
\usepackage{arydshln}
\usepackage{hyperref}
\hypersetup{colorlinks,allcolors=blue}

\makeatletter
\renewcommand*\aa@pageof{, page \thepage{} of \pageref*{LastPage}}
\makeatother
\begin{document}

\title{Gravity darkening and tidally perturbed stellar pulsation in the misaligned exoplanet system WASP-33}

   \author{Sz. K\'alm\'an
            \inst{1}$^,$\inst{2}$^,$\inst{3}
            \and
         A. B\'okon          
            \inst{2}$^,$\inst{4}$^,$ \inst{6}           
          \and
          A. Derekas \inst{2}$^,$\inst{5}$^,$\inst{6}   
          \and
         Gy. M. Szab\'o\inst{2}$^,$ \inst{5}$^,$\inst{6}
         \and
        V. Heged\H{u}s \inst{6}$^,$\inst{7}
         \and
         K. Nagy \inst{8}
          }

   \institute{ 
   Konkoly Observatory, Research Centre for Astronomy and Earth Sciences, Eötvös Loránd Research Network (ELKH), Konkoly-Thege Miklós út 15–17, 1121 Budapest, Hungary
   \and  
   MTA-ELTE Exoplanet Research Group, Szent Imre h. u. 112, 9700 Szombathely, Hungary       
   \and
    ELTE E{\"o}tv{\"o}s Lor\'and University, Doctoral School of Physics,   Pázmány Péter sétány 1/A,1117 Budapest, Hungary
    \and  
  University of Szeged, Department of Optics and Quantum Electronics, Institute of Physics,  D\'om t\'er 9,6720 Szeged, Hungary
  \and        
   ELTE E{\"o}tv{\"o}s Lor\'and University, Gothard Astrophysical Observatory, Szent Imre h. u. 112, 9700 Szombathely, Hungary
   \and        MTA-ELTE  Lend{\"u}let "Momentum" Milky Way Research Group, Hungary
    \and University of Szeged, Institute of Physics,  D\'om tér 9, 6720 Szeged, Hungary
    \and University of Szeged, Doctoral School of Physics,  D\'om tér 9, 6720 Szeged, Hungary
  }

   \date{Recieved ....; accepted ....}

 
  \abstract
   {}
   {WASP-33 is one of the few $\delta$ Sct stars with a known planetary companion. By analyzing the stellar oscillations, we search for possible star-planet interactions in the pattern of the pulsation.}
   {We made use of the Transit and Light Curve Modeller (TLCM) to solve the light curve from the Transiting Exoplanet Survey Satellite (TESS). We include gravity darkening into our analysis.}
   {The stellar oscillation pattern of WASP-33 clearly shows signs of several tidally perturbed modes. We find that there are peaks in the frequency spectrum that are at or near the $3$rd, $12$th and $25$th orbital harmonics ($f_{orb} \sim 0.82$ d$^{-1}$). Also, there is a prominent overabundance of pulsational frequencies rightwards of the orbital harmonics, being characteristic of a tidally perturbed stellar pulsation, which is an outcome of star-planet interactions in the misaligned system. There are peaks in both the $\delta$ Sct and $\gamma$ Dor ranges of the Fourier spectrum, implying that WASP-33 is a $\gamma$ Dor -- $\delta$ Sct hybrid pulsator. The transit light curves are best fitted by a gravity darkened stellar model, and the planet parameters are consistent with earlier determinations.}
   {}

\titlerunning{WASP-33b}
   \maketitle

\section{Introduction}

The known $\delta$ Sct sample from the \verb|Kepler| \citep{2010Sci...327..977B} field has been explored by \cite{2021AJ....162..204H}, who found only $20$ planet candidates in total, shedding light on a notable lack of planetary companions to such stars. This remains true even in the era of 
space-based transiting exoplanet photometry with satellites such as the Transiting Exoplanet Survey Satellite (\verb|TESS|, \citealt{2015JATIS...1a4003R}) and the CHaraterizing ExOplanet Satellite (\verb|CHEOPS|, \citealt{2021ExA....51..109B}), there is a notable lack of planets orbiting $\delta$ Scuti stars. \cite{2021AJ....162..204H} have explored the known $\delta$ Sct sample from the \verb|Kepler| \citep{2010Sci...327..977B} field, and found $20$ planet candidates in total. 

The planetary companion of WASP-33 is a hot Jupiter ($P \sim 1.219867 \pm 4.2\cdot 10^{-5}$ days \citep{2020A&A...639A..34V}, $M_P \sim 2.10 \pm 0.14 M_J$, $R_P \sim 1.59 \pm 0.07 R_J$ \citep{2019AJ....158...39C}). WASP-33b \citep{2006MNRAS.372.1117C} was the first confirmed planet orbiting a $\delta$ Sct host star \citep{2011A&A...526L..10H}, and it has been analyzed thoroughly including Doppler tomography and radial velocity measurements \citep{2010MNRAS.407..507C, 2020PASJ...72...19W, 2021A&A...653A.104B}  as well as ground-based photometry \citep{2014A&A...561A..48V}. Using the TESS light curve, \cite{2020A&A...639A..34V} improved the precision of the known transit parameters. \cite{2021arXiv210903250D} argued that after substracting the pulsation from the TESS light curve, an asymmetry emerges during the transits that is characteristic of a gravity darkened stellar disk.


In this paper we re-examine the TESS light curve of WASP-33, including gravity darkening into our analysis. We also investigate the possibility of star-planet interactions. This paper is structured as follows. In Sect. \ref{sec:lca} we describe our model for the transit in both the gravity darkened and non-gravity darkened cases. In Sect. \ref{sec:res} we re-analyze the TESS light curve of the WASP-33 system using the Transit and Light Curve Modeller (TLCM, \citealt{2020MNRAS.496.4442C}) without pre-whitening the data. We make use of the wavelet-formulation built into TLCM to handle the pulsations and the transit light curve simultaneously. We also reconstruct the Fourier spectra of the stellar oscillations from the fitted light curve, which allowed us to investigate the possibility of planet-star interactions.

\section{Light curve analysis} \label{sec:lca}

\begin{figure*}
    \centering
    \includegraphics[width=\textwidth]{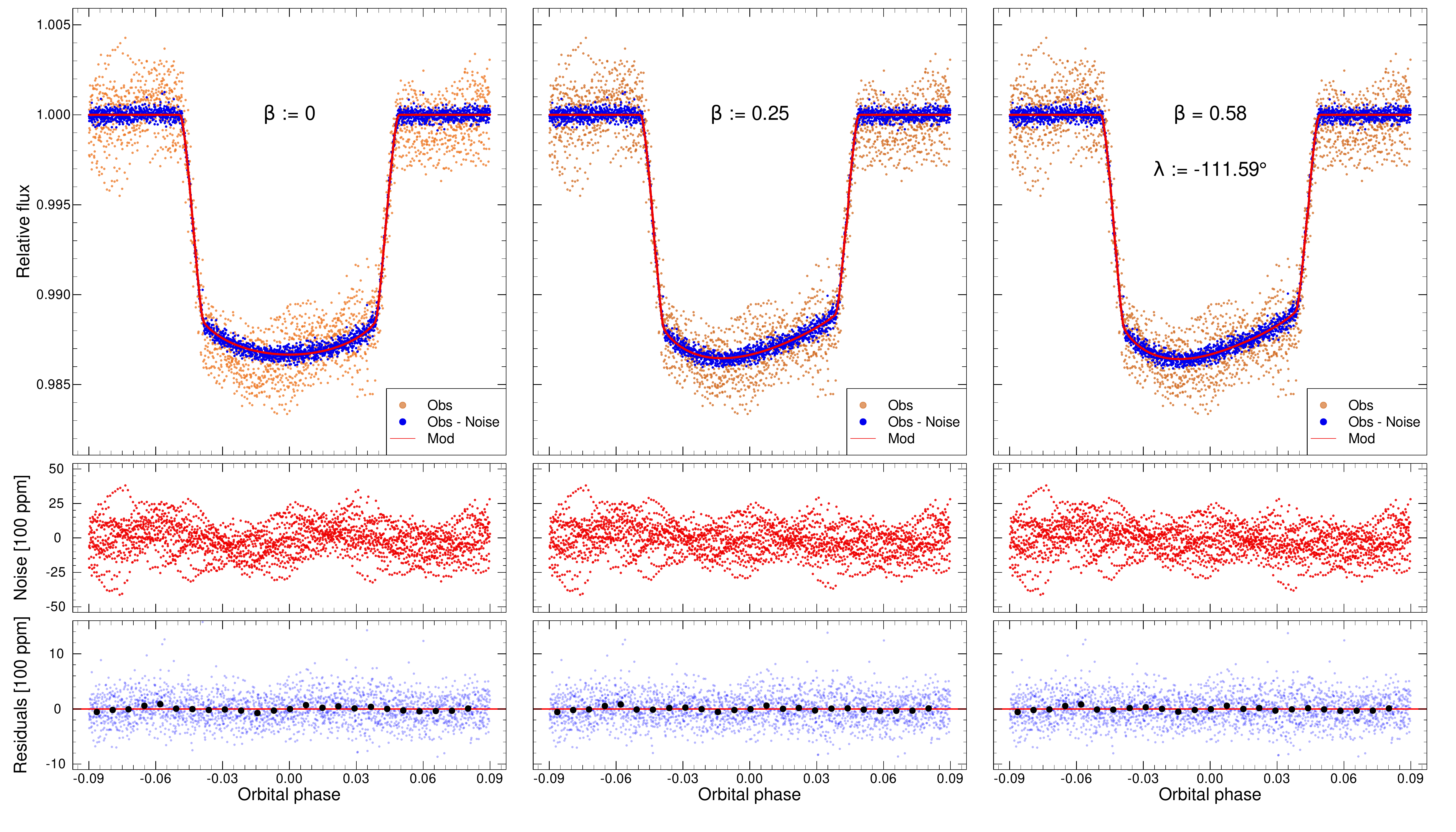}
    \caption{Resulting light curve solutions for the three considered scenarios: $\beta$ fixed to $0$ (left column), $\beta$ set to $0.25$ (middle column) and $\lambda$ set to $-111.59^\circ$ (right column). The top panels show the observed light curve (orange dots), the observed light curves with the noise removed (blue dots) and the transit model (solid red line). The fitted correlated noise (see text for details) is shown in the middle row, while the residuals where the noise and the transit model are subtracted from the observations are shown in the bottom row (blue dots), with bins of $100$ data points (the error bars for the binned points are not shown as they are too narrow). Note that due to the nature of our analysis, the observed light curve (orange points, top panel) is decomposed into the transit model (solid red line, top panel), the correlated noise (middle panel) and the white noise (blue dots, bottom panel).}
    \label{fig:lightcurves}
\end{figure*}

\subsection{Light curve preparation}

The WASP-33 system has been observed in TESS\footnote{The data is available from the Mikulski Archive for Space Telescopes: \url{https://mast.stsci.edu/portal/Mashup/Clients/Mast/Portal.html}} Sector 18. Using the \verb|lightkurve| python package \citep{2018ascl.soft12013L}, we downloaded the PDCSAP data produced by the automated pipeline of the mission. We removed the first $781$ points due to an artifact that could bias the analysis of the first transit event and detrended the light curve. We were eventually left with a dataset of $14028$ measurements consisting of TESS Short Cadence observations, that included 16 transits of WASP-33b. 

\subsection{Light curve solutions} \label{lsc}

\begin{table*}
\caption{Parameters derived from the three considered cases. Bold numbers mark the fixed values for the given parameters.}
\label{tab:params}
\centering
\begin{tabular}{ l c c c }
\hline
\hline
Parameter & {No gravity darkening} & {Fixed gravity darkening exponent} & {Fixed projected spin orbit angle} \\
\hline
$a/R_S$ & $3.651 \pm 0.033$ &  $3.631 \pm 0.045$ & $3.647 \pm 0.034$\\ 
$b$ & $0.132 \pm 0.030$ &  $0.177 \pm 0.039$ & $0.141 \pm 0.030$\\ 
$u_+$ & $0.255 \pm 0.056$ &  $0.243 \pm 0.081$ & $0.310 \pm 0.068$\\ 
$u_-$ & $0.226 \pm 0.174$ &  $0.208 \pm 0.287$ & $0.442 \pm 0.242$\\ 
$R_P / R_S$ & $0.1121 \pm 0.0007$ &  $0.1098 \pm 0.0012$ & $0.1101 \pm 0.0007$\\
$\sigma_r$ [100 ppm] & $434.141 \pm 4.129$ &  $432.885 \pm 6.833$ & $432.824 \pm 4.130$\\ 
$\sigma_w$ [100 ppm] & $4.337 \pm 0.053$ &  $4.331 \pm 0.098$ & $4.331 \pm 0.050$\\ 
$t_C$ [BJD-$2458792$] & $0.63407 \pm 0.00012$ &  $0.63400 \pm 0.00018$  & $0.63402 \pm 0.00011$ \\ 
$P$ [days] & $1.2198635 \pm 0.0000068$  &  $1.2198637 \pm 0.0000103$  & $1.2198636 \pm 0.0000085$ \\
$i_\star$ [$^\circ$] & \multicolumn{1}{c}{\textbf{0}} &  $43.54 \pm 4.99$ & $65.26 \pm 8.72$\\ 
$\lambda$ [$^\circ$] & \multicolumn{1}{c}{\textbf{0}} &  $-94.9 \pm 21.4$ & \multicolumn{1}{c}{\textbf{-111.59}} \\ 
$\beta$ & \multicolumn{1}{c}{\textbf{0}} &  \multicolumn{1}{c}{\textbf{0.25}} & $0.58 \pm 0.20$\\ 
$RSS $ & \multicolumn{1}{c}{0.02669462} & \multicolumn{1}{c}{0.02649666} & \multicolumn{1}{c}{0.02651569}\\
$BIC$ & \multicolumn{1}{c}{-184711.4} & \multicolumn{1}{c}{-184787.2} & \multicolumn{1}{c}{-184777.1}\\
$AIC$ & \multicolumn{1}{c}{-50813.6} & \multicolumn{1}{c}{-50912.0} & \multicolumn{1}{c}{-50901.9}\\

\hline
\end{tabular}
\end{table*}

\begin{table*}
    \centering
    \caption{Comparison of our derived parameters to the non-gravity darkened model of \cite{2020A&A...639A..34V} and the gravity darkened model of \cite{2021arXiv210903250D} -- of the same TESS light curve. Bold numbers mark the fixed values of the given parameters.}
    \label{tab:comp}
    \begin{tabular}{l c c c}
    \hline
    \hline
    Parameter & {This work} & {Accepted value} & Reference \\
    \hline
       $a/R_S$  &  $3.631 \pm 0.045$ &  $3.614 \pm 0.009$ & \multirow{7}{*}{\text{\cite{2020A&A...639A..34V}}} \\
       $i$ [$^\circ$] &  $87.20 \pm 1.25$ &  $88.01 \pm 0.03$ &\\
       $R_P/R_S$ &  $0.1098 \pm 0.0012$ &  $0.1071 \pm 0.0002$ & \\
       $u_+$ &  $0.243 \pm 0.081$ &  \multicolumn{1}{c}{\textbf{0.498}} & \\
       $u_-$ &  $0.208 \pm 0.287$ &  \multicolumn{1}{c}{\textbf{-0.006}} & \\
       $P$ [days] &  $1.2198637 \pm 0.0000103$ &  $1.2198681 \pm 0.0000042$  &  \\
       $t_C$ [BJD-2458792] &  $0.63400 \pm 0.00018$ &  $0.63403 \pm 0.00009$ &  \\  
        \rule{0pt}{3ex}$i_\star$ [$^\circ$] &  $43.54 \pm 4.99$ & \multicolumn{1}{c}{$69.8^{+4.0}_{-3.2}$} & \multirow{4}{*}{\text{\cite{2021arXiv210903250D}}} \\
       $\lambda$ [$^\circ$] &  $-94.9 \pm 21.4$ &  \multicolumn{1}{c}{$-109.0^{+20.2}_{-17.6}$} & \\
       $\varphi$ [$^\circ$] & $91.3\pm 18.5$ & \multicolumn{1}{c}{$108.3^{+19.0}_{-20.2}$} & \\
       $\beta$  &  \multicolumn{1}{c}{\textbf{0.25}} &  \multicolumn{1}{c}{\textbf{0.23}} & \\
        \hline
    \end{tabular}
\end{table*}

Stellar oscillations and any possible instrumental effects severely distort the transit light curve of WASP-33b. In order to compensate for these effects, we employed the wavelet-based routines of \cite{2009ApJ...704...51C}, built into TLCM. These describe the time-correlated noise (i.e. pulsation and systematics) in terms of two parameters: $\sigma_r$ for the red component and $\sigma_w$ for the white component (see \citealt{2009ApJ...704...51C} for details). \cite{2021arXiv210811822C} proved the validity of this approach, while \cite{powIII} completed an extensive test for the consistency of transit parameter estimation using TLCM. This code allows the simultaneous fitting of the noise and the transit. The transit modeling is carried out via the analytic model of \cite{2002ApJ...580L.171M} which is described by the ratio of the planetary and stellar radii, $R_P/R_S$, the scaled semi-major axis, $a/R_S$, the impact parameter, $b = a/R_S \cos i$ (where $i$ is the orbital inclination relative to the line of sight), and the time of midtransit, $t_C$.

Gravity darkening of the host star causes asymmetric transit light curves \citep{2009ApJ...705..683B}. This phenomenon, as implemented by TLCM, is described by three parameters \citep{2021arXiv210811822C}: the gravity darkening exponent, $\beta$, the inclination of the stellar rotational axis (measured from the line of sight), $i_\star$, and the projected spin orbit angle $\lambda$. In order to perform a thorough analysis of the WASP-33b light curve, we considered three different scenarios: (i) no gravity darkening (despite the rapid rotation of WASP-33 at $v\sin i_\star = 86.5$ km s$^{-1}$, \citep{2015ApJ...810L..23J}), (ii) gravity darkening exponent fixed to it's nominal value $0.25$ \citep{1924MNRAS..84..684V} while $i_\star$ and $\lambda$ are free and (iii) $\lambda$ fixed to $-111.59^\circ$ \citep{2021A&A...653A.104B} with $i_\star$ and $\beta$ as free parameters. In the latter two cases, knowing the orbital inclination $i$, as well as $i_\star$ and $\lambda$ allows for the calculation of the true spin orbit angle $\varphi$ via \citep{2009ApJ...696.1230F}:
\begin{equation}
    \varphi = \arccos \left( \cos i_\star \cos i + \sin i_\star \sin i \cos \lambda \right).
\end{equation}
We note that without RV data, it is not possible to fit for $\beta$, $i_\star$ and $\lambda$ at the same time using the approach embedded into TLCM.

We made use of a quadratic limb darkening formula (described by the limb darkening coefficients $u_1$ and $u_2$), but as theoretical limb darkening coefficients are calculated for spherically symmetric and static stars (and WASP-33 with its rapid rotation and pulsation is neither), we left $u_{\pm} = u_1 \pm u_2$ as free parameters. We used the system parameters determined by \cite{2020A&A...639A..34V} as starting values for the fitting processes. We set the third light to $0.024$ \citep{2020A&A...639A..34V}, while fixing $T_{eff} = 7430 \pm 100$ K \citep{2010MNRAS.407..507C} and $\log g = 4.25 \pm 0.10$ \citep{2019AJ....158..138S}. We assumed a circular orbit and set the mass ratio to $q = M_P/ M_S = 0.001335$ \citep{2019AJ....158...39C}.

\section{Results}\label{sec:res}

\begin{figure*}
    \centering
    \includegraphics[width=\textwidth]{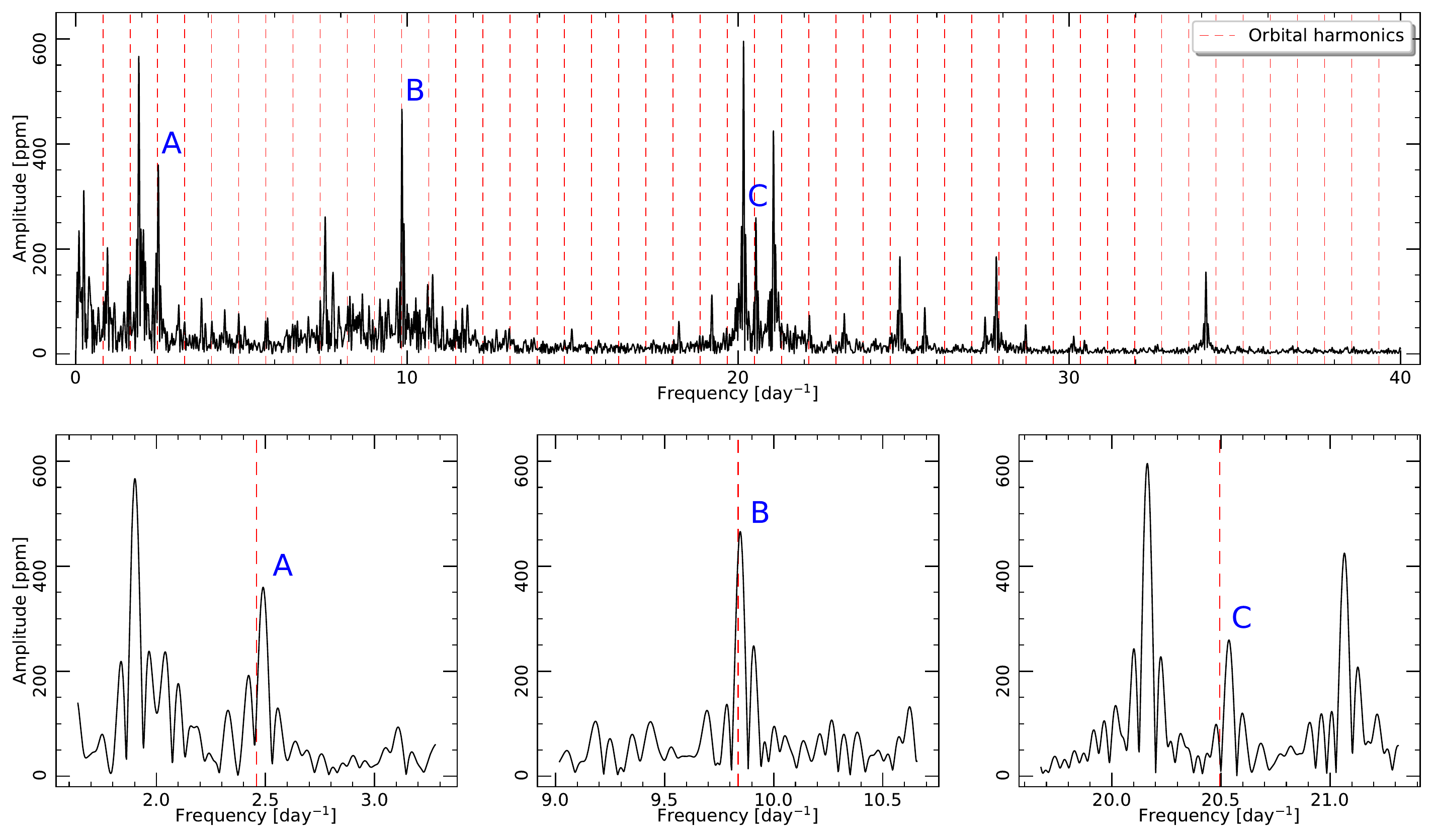}
    \caption{Computed Fourier spectrum (top panel, black), the orbital frequency and the first 47 orbital harmonics (top panel, red dashed lines) that cover the range of frequencies $< 40$ days$^{-1}$, above which no significant peaks are found. The bottom row shows the $3$rd, $12$th and $25$th orbital harmonics (red dashed lines) and the zoomed-in parts of the spectrum that correspond to these frequencies. In both rows, $A$, $B$ and $C$ denote the peaks that are closest to these harmonics.}
    \label{fig:spec}
\end{figure*}

\begin{figure}
    \centering
  \includegraphics[width = .5\textwidth]{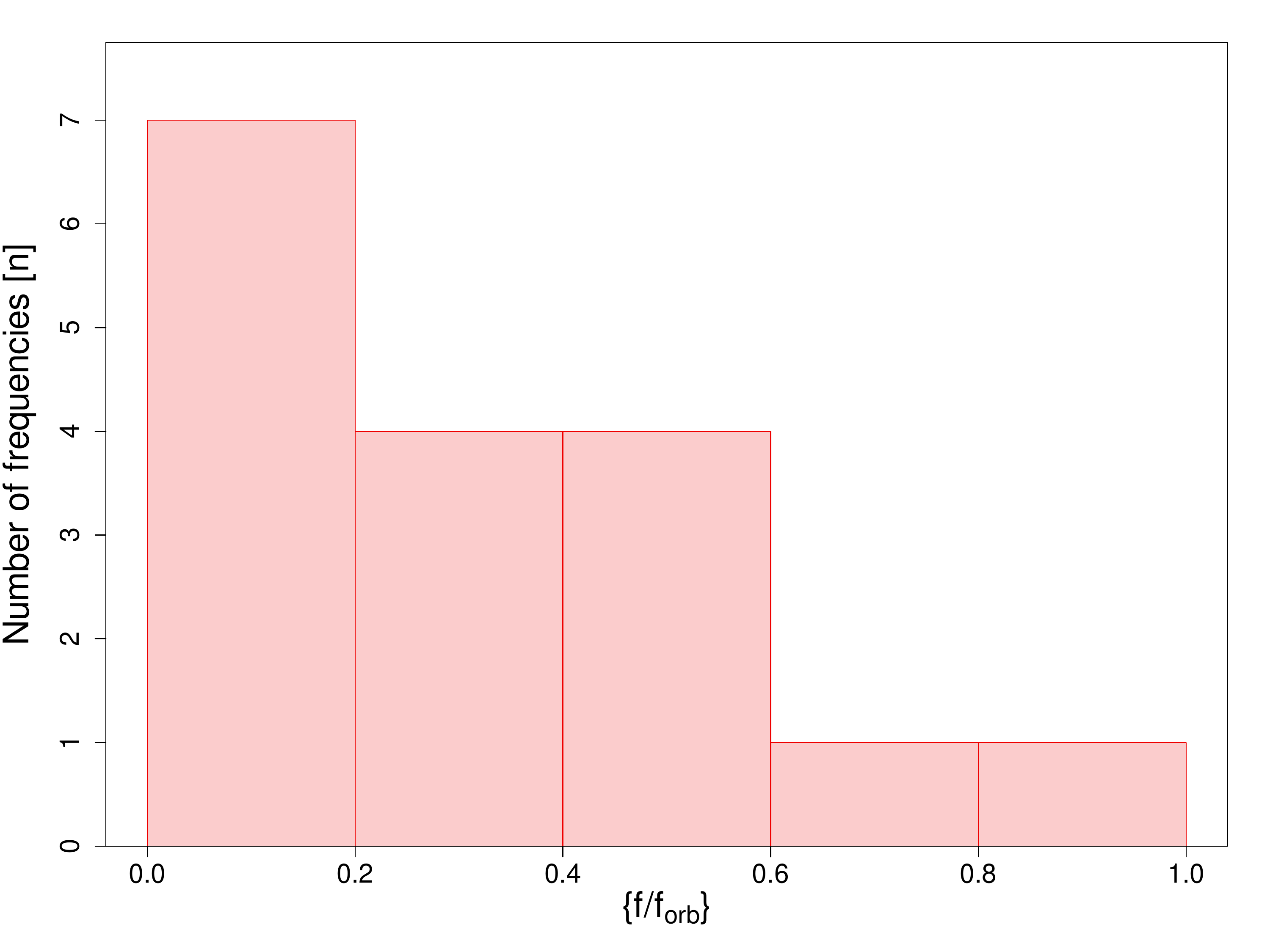} \\
  \includegraphics[width = .5\textwidth]{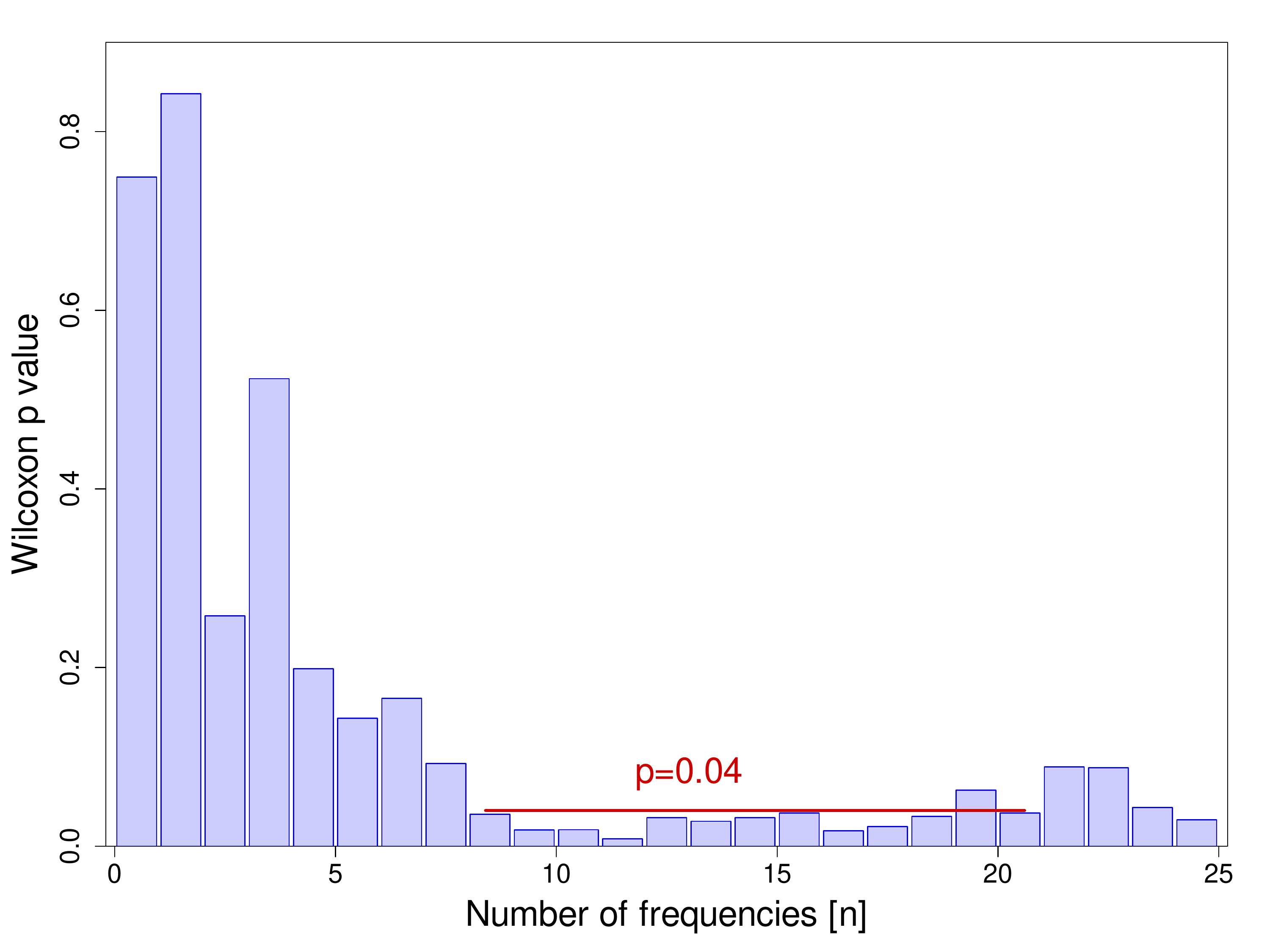}    \caption{Distribution of the fractional frequencies of the $17$ highest amplitude components (upper panel) and the Wilcoxon-$p$ values of the fractional part of the $n$ highest amplitude frequencies compared to a uniform distribution (lower panel).}
    \label{fig:ffraction}
\end{figure}

\subsection{System parameters from TLCM}

The results of the fitted and phase-folded light curves of the three different cases outlined in Sect. \ref{lsc} are shown in Fig. \ref{fig:lightcurves}. When $\beta = 0$ is assumed, the correlated noise (i.e. pulsation + systematics) agrees well with the residuals that one would expect from a planet transiting a gravity darkened star. This is because the simultaneous fitting of the noise and the transit signal leads to the asymmetry being treated by the wavelet-formalism as part of the time-correlated noise, as seen in the middle and bottom panels of Fig. \ref{fig:lightcurves}. 

To decide which model describes the system better, we computed the Bayesian and Akaike Information Criteria (BIC and AIC) for the transit model \textit{only} (without the noise) as $BIC = N_{obs} \cdot \ln \left( RSS / N_{obs} \right) + N_{params} \cdot \ln N_{obs}$ and $AIC = 2 \cdot N_{params} + N_{obs} \cdot \ln RSS  + const.$, where the constant term is the same for all three scenarios, thus can be disregarded. Here $N_{obs}$ is the number of measurements used during the fit, $N_{params}$ is the number of parameters included in the model and $RSS$ is the residuals sum of squares. The lower $BIC$ and $AIC$ numbers are diagnostic of more plausible models.

The resulting parameter sets for the three cases are shown in Tab. \ref{tab:params}. Almost all parameters, with the obvious exception of the three variables characterizing the gravity darkening ($i_\star$, $\lambda$, and $\beta$), are consistent with each other within the estimated uncertainty range. The discrepancies in $b$ between cases (i) and (ii) ($\beta = 0$ and $\beta = 0.25$) are within $3\sigma$ of each other. This means that $R_P/R_S$ is the only parameter where the fitted value in the two cases are significantly different (similar observations can be made when comparing our findings to the literature, see below). The transit light curve of the first case, where gravity darkening assumed to be negligible, is a function of $7$ parameters, while the other two are determined by $10$ parameters. After calculating the Bayesian and Akaike Information Criteria ($BIC$ and $AIC$), we can infer that the $\beta = 0.25$ case is the most probable out of the three, as it is $\Delta BIC =75.8$, $\Delta AIC = 98.4$, $\Delta BIC = 10.1$, and $\Delta AIC = 10.1$ lower than the $\beta = 0$ and $\lambda = -111.59^\circ$ cases, respectively. Thus we confirm the presence of gravity darkening in the WASP-33b system first noted by \cite{2021arXiv210903250D}, and adopt this model.   

Comparing the parameters of our adopted model to those from the literature (Tab. \ref{tab:comp}), we find good agreements with the no gravity darkening model of \cite{2020A&A...639A..34V} for $a/R_S$, $i$, $P$, $t_C$ and (because of the large uncertainty) $u_-$. There is a $3.1 \sigma$ difference between our fitted value of $u_+$ compared to the same parameter from \cite{2020A&A...639A..34V}, but we note that the authors of that paper used theoretical limb darkening coefficients. Of more interest is the ratio of the planetary and stellar radii as our value of $0.1098 \pm 0.0012$ is in good agreement with \cite{2021arXiv210903250D} ($0.1088 \pm 0.0003$), but is is significantly different from the $0.1071 \pm 0.0002$ derived by \cite{2011A&A...526L..10H}. Comparing the results of $i_\star$, $\lambda$ and $\varphi$ to those from \cite{2021arXiv210903250D} (who also fixed the value of the gravity darkening exponent), we find good agreements for the projected and true spin orbit angles, while the stellar inclinations are inconsistent with each other.

The resulting projected spin orbit angle of $\lambda = -94.9^\circ \pm 21.4^\circ$ is in good agreement with the retrograde motion discovered by \cite{2010MNRAS.407..507C}. However, because of the wide uncertainty range, we are unable to consider the nodal precession observed by e.g. \cite{2015ApJ...810L..23J}, \cite{2020PASJ...72...19W} and \cite{2021A&A...653A.104B}.

We also note that setting the gravity darkening exponent as a free parameter (as in our (iii) case) yields $\beta = 0.58 \ 0.20$ which is significantly larger than the nominal value of $0.25$, however, it fits in reasonably well with the trend established by \cite{2003A&A...402..667D} and \cite{2006A&A...445..291D}.


\subsection{Stellar oscillations}

The wavelet-formulation built into TLCM fits correlated noise simultaneously with the transits, which allowed us to analyze the stellar oscillations without masking the transits. We made use of  \verb|Period04| \citep{2005CoAst.146...53L} to calculate the spectrum of the identified correlated noise from the adopted solution (where $\beta = 0.25$, Fig. \ref{fig:spec}). Upon a visual inspection, this spectrum looks identical to the one published by \cite{2020A&A...639A..34V}. There are peaks in both the $\delta$ Sct and $\gamma$ Dor ranges of the spectrum, which implies that WASP-33 is a $\gamma$ Dor -- $\delta$ Sct hybrid pulsator \citep{2020A&A...639A..34V}.

We identified three pulsational frequencies exceeding S/N $>8$ which are very close to one of the orbital harmonics (especially, the $3$rd,  $12$th, and $25$th orbital harmonics) in the frequency spectrum. The properties of these peaks are summarized in Tab. \ref{tab:harmonics}. A magnification to the environment of these frequencies and the closest orbital harmonics are shown in the bottom row of Fig. \ref{fig:spec}. Indeed, the pulsational frequencies are close to the orbital harmonics, and while there is no exact coincidence, the pulsational frequencies are observed to be shifted a little bit rightwards (towards higher frequencies). The rightward shifted pulsational frequencies are known as perturbed frequencies and they are similarly interpreted in the case of
 V453 Cyg \citep{2020MNRAS.497L..19S}, V1031 Ori \citep{2021PASJ...73..809L} and RS Cha \citep{2021A&A...645A.119S}. In these systems, the perturbations have been identified in \textit{p} modes alone. V453 Cyg, V1031 Ori, and RS Cha are all close binaries with circular orbits that possibly exhibit spin-orbit misalignment, meaning that WASP-33 is a close analog to them. Two of the three frequencies (Fig. \ref{fig:spec} bottom row) where the tidal perturbations are the most prominent (near the $12$th and $25$th orbital harmonics) are likely also pressure driven modes, which fits in well with the preferred interpretation for this phenomenon. However, \cite{2010MNRAS.407..507C} found from spectroscopic analysis that WASP-33 exhibits \textit{g} mode pulsations, too. As the peak in the Fourier spectrum near the $3$rd harmonic likely is caused by \textit{g} mode oscillations, we suggest that in the case of WASP-33, uniquely, tidally perturbed oscillations arise in both pressure and gravity driven modes as well.

 
 In order to investigate the rightward-shifted nature of the pulsational frequencies compared to the orbital harmonics, we created a noise model with the same length as the input light curve, with a Gaussian distribution whose standard deviation is equal to the standard deviation of the residuals of the fitted light curve ($\sim 260$ ppm). We injected a sinusoidal signal into this simulated noise model with the same frequency as the orbital harmonic, a randomized phase between $0$ and $2 \pi$, and an amplitude roughly equal to the height of the respective peaks in the spectrum ($340$, $440$ and $240$ ppm). We calculated the Fourier spectrum of the resulting sinusoidal light curve, extracted the frequency of its single peak, then repeated the process 1000 times. We have found that the widths of the distributions constructed from the bootstrapped frequencies (not shown here) thus determined are negligible  compared to the widths of the peaks resulting from the stellar pulsations (Fig. \ref{fig:spec}, lower row) at $0.0014$, $0.0011$ and $0.0020$ d$^{-1}$ compared to $\sim 0.1$ d$^{-1}$. This confirms that the relative positions of the three inspected harmonics compared to the peaks in the spectrum that they correspond to are not caused by a stochastic phenomenon of the particular noise realisation of the original observation.

The distribution of the frequencies in Fig. \ref{fig:spec} also suggests that the rightward position of the pulsational frequencies in relation to the closest harmonics of the orbital period may extend to a much larger set of the pulsational frequencies than the three most evident examples shown in the lower row of Fig. \ref{fig:spec}. To examine the position of the stellar frequencies in the "frequency comb" of rotational harmonics, we defined the "fractional frequency" belonging to the stellar modes as
\begin{equation}
    \left\{ \frac{f}{f_{orb}} \right\} = \frac{f}{f_{orb}} - \text{int}\left(\frac{f}{f_{orb}} \right),
\end{equation}
where \verb|int| represents the integer part of a number. The histogram of the fractional frequencies is shown in the top panel of Fig. \ref{fig:ffraction}. The distribution consists of the 17 stellar frequencies with the highest amplitudes. The distribution is very heavy towards the small fractional parts, 13 of 17 frequencies belong to the left half of the distribution ($\{f/f_{orb}\} < 0.5$), while only 2 stellar frequencies are observed with $\{f/f_{orb}\} > 0.6$. This asymmetry is not compatible with a stochastic realisation of a uniform distribution. (A uniform distribution would suggest a random set of pulsation frequencies, which are expected if the frequency pattern is not biased/filtered or otherwise modified by the orbital motion of the planet.) We performed a Wilcoxon test \citep{1993stp..book.....L} to compare the distribution of the observed modular frequencies to a uniform distribution. The $p$ value of the Wilcoxon test measures the probability of having "at least as asymmetric" distribution of completely random set of numbers as it was observed in the spectrum of the stellar signal, just due to numerical fluctuations. Thus, if $p$ is small, the observed distribution comes from a uniform distribution with a very low probability. The Wilcoxon $p$ value is plotted in the bottom panel of Fig. \ref{fig:ffraction} as the function of the $n$ number of frequencies included in the test (while the frequencies were sorted according to their amplitude).

We find that $p<0.04$ in the range of $n$ between 8 and 19. Fewer number of frequencies lead to lack of significance due to the low number statistics, while more frequencies probably include an increasing ratio of those low-amplitude frequencies which are not related to the orbital frequency of the planet.

\begin{figure}[!h]
    \centering
    \includegraphics[width = .5\textwidth]{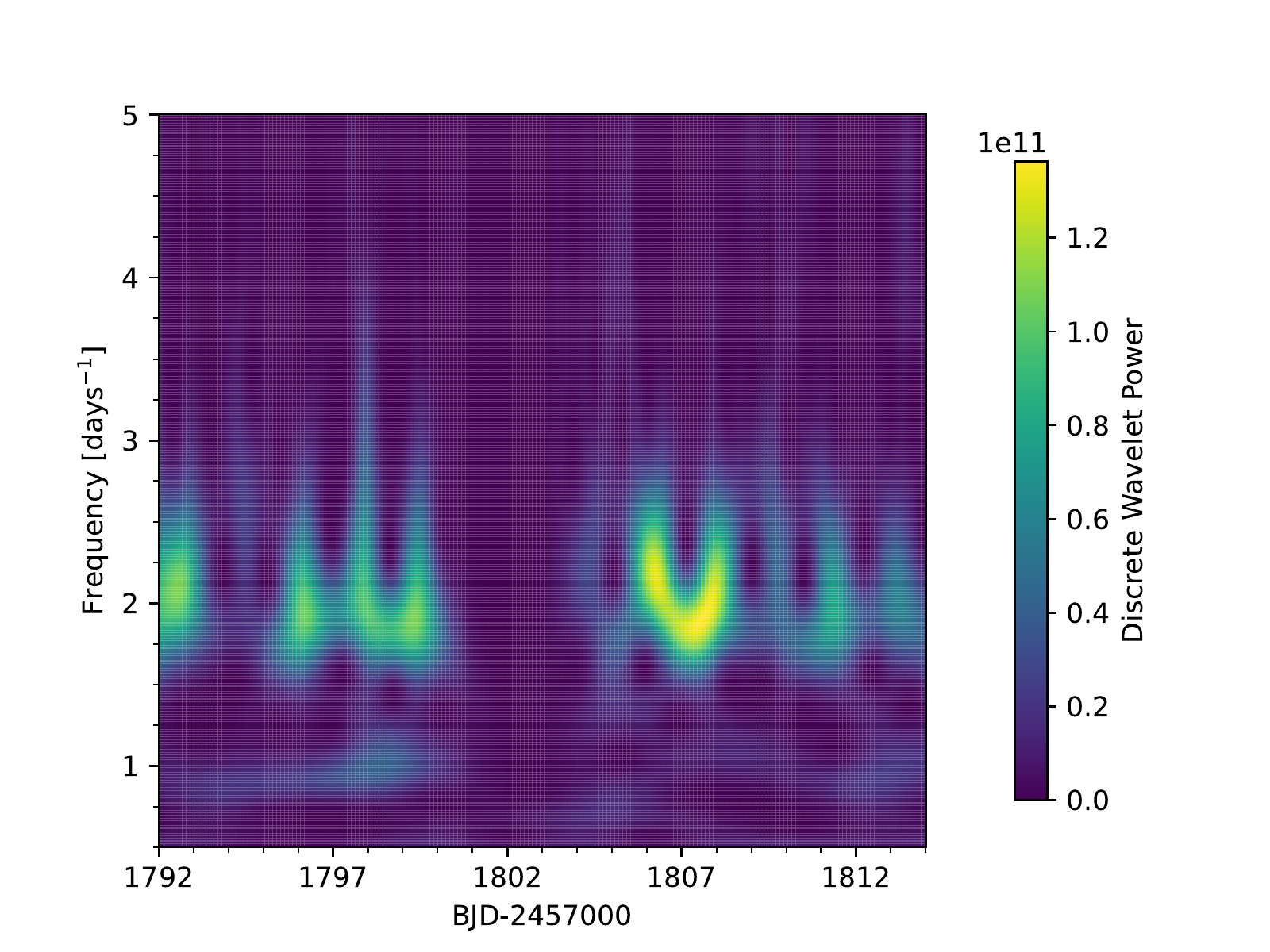}
    \caption{Discrete wavelet map of the identified correlated noise (stellar oscillations $+$ systematics) in the frequency range of $0.5$ -- $5$ d$^{-1}$. The gap between $\sim 1801$ and $1804$ days is the result of the TESS data downlink.}
    \label{fig:wav}
\end{figure}

To reveal the plausibility of mode interactions due to possible hydrodynamical interactions between the modes of stellar pulsation and the tidal forces generated by the planet on the misaligned orbit, we calculated the discrete wavelet transform of the reconstructed stellar signal for frequencies $\leq 5$ d$^{-1}$, belonging to the first major complex of frequencies towards the long-period end of the frequency distribution (Fig. \ref{fig:spec}). The wavelet map of this frequency region is shown in Fig. \ref{fig:wav}. The wavelet map shows vivid amplitude and frequency modulations in the timescale  $\sim~1.5$--$2$ orbits. 
The degree and the pattern of the detected instability of amplitudes and frequencies in the case of WASP-33 rather suggests a recurrent redistribution of pulsation energy between different modes. Further visual inspection of the wavelet map shows that the oscillations corresponding to the highest two peaks  of the bottom left panel of Fig. \ref{fig:spec} ($\sim 2$ and $\sim 2.5$ d$^{-1}$) are not constantly present. It can also be pointed out that the two frequencies are not present simultaneously, and that the oscillations transfer from one mode to the other regularly. This may be caused by star-planet interactions, as the stellar spin axis and the orbital plane are not aligned. Due to sampling and smearing, the wavelet transform cannot be used in this analysis at higher frequencies.

We note that Fig. \ref{fig:spec} shows the frequency reconstruction with the fixed $\beta=0.25$ parameter during the fit, belonging to the removal of a transit model with standard gravity darkening before this analysis. We repeated the same analysis as above for the two other planet solutions with different handling of the gravity darkening (no gravity darkening, free $\beta$: left and right columns of Fig. \ref{fig:lightcurves}), and we got practically the same result. Only a slight change in the amplitudes of the peaks can be observed, the frequencies themselves are not affected, and they are always rightwards to the closest orbital harmonics.

The process behind this frequency pattern should be more complex than in the case of ``heartbeat stars'' (e.g. \citealt{2013MNRAS.434..925H, 2016MNRAS.463.1199H})
because in the case of WASP-33 the pattern does not strictly follow the orbital period. But together with the distribution of the fractional frequencies (Fig. \ref{fig:ffraction}), this is a strong evidence for the tidally perturbed frequency pattern of the host star. It is also different from tidally excited oscillations, as in those cases the eccentric orbits result in a strong coincidence of the orbital harmonics and the peaks in the Fourier spectrum (see e.g. \cite{2017MNRAS.472.1538F, 2021FrASS...8...67G}).

\begin{table}
\centering
\caption{Frequency, amplitude and S/N of the peaks that are found to be near an orbital harmonic.} \label{tab:harmonics}
\begin{tabular}{c c c c c}
\hline
\hline
$f$ [d$^{-1}$] & Amplitude [ppm] & $n$ & $f-n/P$ [d$^{-1}$] & S/N\\
\hline
$2.49043$ &  $343.620$  & $3$ & $0.03113$ & $12.750$ \\ 
$9.84947$ &  $446.224$  & $12$ & $0.01230$ & $12.725$ \\ 
$20.53765$ &  $237.176$  & $25$ & $0.04355$ & $8.835$ \\ 
\hline
\end{tabular}
\end{table}

\section{Summary}

We have performed a re-analysis of the light curve of WASP-33 observed by TESS in Sector 18. Our analysis is unique in the way that we did not perform a pre-whitening of the light curve before modeling the transit. Instead, we used the wavelet formalism built into the TLCM code to fit the stellar pulsations, instrumental noise and the transit signal of WASP-33b simultaneously. 

To test for gravity darkening, we considered three scenarios: no gravity darkening; gravity darkening exponent $\beta$ fixed to $0.25$ and $\beta$ set as a free parameter of the fit. We have found that the $\beta = 0.25$ case describes the light curve best, thus we confirm the findings of \cite{2021arXiv210903250D}. 

We also investigated the Fourier spectrum of the stellar oscillations reconstructed by TLCM. There are three peaks in the spectrum (with an S/N $> 8$) that are at or near an orbital harmonic ($3$rd, $12$th and $25$th). The distribution of the $17$ frequencies with the highest amplitudes in relation to the orbital harmonics let us to conclude that the frequencies in the Fourier spectrum tend to be shifted rightward compared to the harmonics. Using a two sampled Wilcoxon test, we compared their distribution to a uniform distribution of random numbers, and found that this rightward shifted nature is highly unlikely to be caused by a stochastic phenomenon. We suggest that this effect is caused by tidal perturbations of the planetary companion. This would also make WASP-33 the first system where tidal perturbations of the pulsational modes are caused by a sub-stellar companion. The significant changes of frequency and amplitude in the $f < 5$ days$^{-1}$ region of the spectrum seen on the wavelet map (Fig. \ref{fig:wav}) are also diagnostic of planet-star interactions. While the possibility of these interactions has been noted previously by \cite{2010MNRAS.407..507C} and \cite{2011A&A...526L..10H} (especially at the frequency of $21.311$ d$^{-1}$, a peak altogether missing from our spectrum), our analysis gives detailed, statistical proofs of the stellar oscillations caused by tidal forces.   We also suggest that the misalignment of the stellar rotational axis and the planetary orbit is causing the perturbations of the stellar pulsational modes.



\begin{acknowledgements}
We acknowledge the support of the Hungarian National Research, Development and Innovation Office (NKFIH) grant K-125015, a PRODEX Experiment Agreement No. 4000137122 between the ELTE E\"otv\"os Lor\'and University and the European Space Agency (ESA-D/SCI-LE-2021-0025), the Lend\"ulet LP2018-7/2021 grant of the Hungarian Academy of Science and the support of the city of Szombathely. Prepared with the professional support of the Doctoral Student Scholarship Program of the Co-operative Doctoral Program of the Ministry of Innovation and Technology financed from the National Research, Development and Innovation Fund.
\end{acknowledgements}

\bibliographystyle{aa}
\bibliography{refs}
\end{document}